\documentclass[nopubldata]{lpor2014}
\graphicspath{{figs/}}
\begin{document}
%
\titlefigure{fig-intro}

\abstract{The most commonly observed quantity related to light is its power or equivalently its energy. It can be either measured with a bolometer, a photodiode or estimated with the naked eye. Alternatively people can measure the light impulse or linear momentum. However, linear momentum is characterized by its transfer to matter, and its precise value is most of the time of little use. Energy and linear momentum are linked and can be deduced from each other, from a theoretical point of view. Because the linear momentum measurement is more difficult, energy is the most often measured quantity. In every physical process, angular momentum, like energy and linear momentum is conserved. However, it is independent and can't be deduced from the energy or the linear momentum. It can only be estimated via its transfer to matter using a torque observation. Nevertheless, experimentally, the torque is found to be proportional to the optical power. This leads to a need of a quantum interpretation of the optical field in terms of photons. Clear experimental evidences and consequences are presented here and debated.}

\title{Energy, linear and angular momentum of light: what do we measure?}
%
\titlerunning{Light: what do we measure?}
\author{Olivier EMILE\inst{1,*}, Janine EMILE\inst{2}}
%
\authorrunning{O. Emile and J. Emile}
%
\institute{Universit\'{e}  Rennes 1, 35042 Rennes cedex, France
\and
IPR, UMR CNRS 6251, Universit\'{e}  Rennes 1, 35042 Rennes cedex, France}
%
\mail{\email{olivier.emile@univ-rennes1.fr}}
%
%
\maketitle

\section{Introduction}

In everyday life, people are used to the notion of light detection and to the manipulation of light detectors. This detection can be performed  either with the naked eye or with dedicated sensors. Nevertheless, light characterization and more generally electromagnetic radiation measurement is a problem of light/matter interaction. More specifically certain characteristics of the electromagnetic field must be transferred to electrons. Up to few GigaHertz,  \emph{\textbf{free}} electrons of matter, like in metals for antennas, oscillate directly in response to the electromagnetic field solicitation. At higher frequencies, due to their finite mass, and because of energy and momentum conservations, the free electrons can't respond anymore. In the hundred of GigaHertz range, graphen or carbon nanotubes, that can be considered as two dimensional gas of massless carriers, respond to an electromagnetic excitation \cite{Novoselov2005,Geim2007,Liao2010,Zhong2017}. It has to be noted that in the Drude model for metals \cite{Drude1900}, free electrons should respond up to 50 THz. However, for simple real metals other features such as band structure play an important role \cite{Dressel2006} and such behavior of electrons is not experimentally observed. Curiously, in an even higher frequency range, in the optical domain, the \emph{\textbf{bounded}} electrons of matter (atoms, molecules, or solids) are responsible for light matter interaction detection. These electrons make transitions between different levels or bands. 

On the other hand, concerning electromagnetic fields characterization or observation, the fields are themselves only abstractions that can't be directly measured. Quoting Dyson \cite{Dyson1999} 
 \begin{quote}
"The modern view of the world that emerged from Maxwell's theory is a world with two layers. The first layer, the layer of the fundamental constituents of the world, consists of fields satisfying simple linear equations. The second layer, the layer of the things that we can directly touch and measure, consists of mechanical stresses and energies and forces. The two layers are connected, because the quantities in the second layer are quadratic or bilinear combinations of the quantities in the first layer \dots The objects on the first layer, the objects that are truly fundamental, are abstractions not directly accessible to our senses. The objects that we can feel and touch are on the second layer, and their behavior is only determined indirectly by the equations that operate
on the first layer. The two-layer structure of the world implies that the basic processes of nature are hidden from our view.Ó ÒThe unit of electric field-strength is a mathematical abstraction, chosen so that the square of a field strength is equal to an energy-density that can be measured with real instruments... It means that an electric field-strength is an abstract quantity, incommensurable with any quantities that we can measure directly."
\end{quote}

In other words, the electromagnetic fields can only be estimated indirectly from physical observables that are quadratic or bilinear in these fields \cite{Crawford1968,Thide2015}. In telecommunication systems, for example, either in radio or in optics, signals are encoded into electromagnetic fields, modulating the intensity or the frequency of the light beam. Sometimes polarization \cite{Liu2013,ZhangJ2015,Chen2017} or orbital angular momentum \cite{Tamburini2012,Wang2012,Yan2014,Willner2015} are multiplexed, introducing sub-channels in order to increase the channel capacity of a single frequency. One may then wonder what do we finally measure when we introduce a detector to recover the information. The aim of this review is to investigate the quantities we do  actually observe when we perform light detection, in particular when trying to characterize the electromagnetic field. We will wonder which quadratic or bilinear quantities we have access to when we perform an experiment in terms of energy, linear momentum and angular momentum. Besides, since all these quantities are conserved, we will investigate the links between each quantity and explore the consequences concerning these measurements.

To answer such questions the review is organized as follows. After the introduction of few theoretical formulas concerning electromagnetic fields, in section \ref{theory} in a classical approach that follows Maxwell's equations, we will be interested in the energy measurement in section \ref{energy}. Energy is probably the quantity most commonly used to characterize electromagnetic fields in the visible region of the spectrum. The various means to detecting energy we will be discussed: bolometers \ref{bolometer}, photomultipliers  \ref{PM}, photodiodes \ref{SC}, or vision \ref{eye}. These detectors are well known, sometimes already taught at undergraduate level. We will not spend too much time describing them. Section \ref{linear} is devoted to the linear momentum detection, which is also a quadratic quantity in the electromagnetic fields. Its measurement is in principle equivalent to energy measurements, although people are usually less familiar with it and with its underlying concepts. We will focus on the first experimental apparatus dedicated to its mechanical detection \ref{Nichols}, and present other evidence of radiation pressure \ref{solar}. We will then discuss a major controversy about it \ref{abraham}, the so-called Abraham-Minkowsky controversy. Section \ref{am} deals with the angular momentum measurements, either spin angular momentum in sub-section \ref{sam} or orbital angular momentum (sub-section \ref{oam}), that have been hardly considered in the literature. We focus on their specificity and their detection that may sometimes be rather tricky. As for linear momentum, it is mainly from mechanical consequences (rotation) that angular momentum can be detected. We will then discuss the relationships and inter-dependencies between these quantities (section \ref{discussion}) that lead to dramatic consequences for the nature of light, before reaching conclusions. 

\section{Theoretical considerations}
\label{theory}

The aim of this section is to introduce some of the quantities that may be useful in discussing energy, linear momentum, or angular momentum transfer and detection. All these quantities will be treated classically, without introducing any quantization of the electromagnetic fields. They are deduced from Maxwell's equations which form the very basis of classical electromagnetism. These quantities may be found in every book on electromagnetism (see for example \cite{Schwinger1998,Jackson1998,Grifftihs1999,Feynman2005}).

Let us introduce first the energy density $u$. It can be written
\begin{equation}\label{eq1} 
u=\frac{1}{2}(\epsilon_0\mathbf{E}^2+\frac{1}{\mu_0}\mathbf{B}^2),
\end{equation} 	
where $\mathbf{E}$ is the electric field, $\mathbf{B}$ is the magnetic field, $\epsilon_0$ is the vacuum permittivity and $\mu_0$ is the magnetic permeability. When the power of an electromagnetic field is measured, it corresponds to an energy received per second. It is the integral of the energy density over the surface of the detector times the velocity of light, assuming a uniform velocity. The energy is a conserved quantity. It can be stored or transformed into another form such as thermal or chemical energy that can then be transformed into an electrical current. Finally, this current can be detected, as will be discussed in \ref{energy}. 

Concerning the linear momentum or linear impulse, the Poynting vector $\mathbf{S}$ is defined as 
\begin{equation}\label{eq2} 
\mathbf{S}=\frac{1}{\mu_0}\mathbf{E}\otimes\mathbf{B},
\end{equation} 
where $\otimes$ denotes the vector product. The linear momentum density $\mathbf{p}$ (i.e., momentum  per  unit  volume) is related to the Poynting vector by the following relation 
\begin{equation}\label{eq3} 
\mathbf{p}=\mathbf{S}/c^2,
\end{equation}
where $c$  is the velocity of light in vacuum. Linear momentum like energy is a conserved quantity. There is a relation between the density of energy and the Poynting vector called the Poynting theorem. It can be deduced from the Maxwell's equations. It reads
\begin{equation}\label{eq4} 
\frac{\partial u}{\partial t}=-\nabla.\mathbf{S}-\mathbf{j}.\mathbf{E},
\end{equation}
where $\mathbf{j}$ is the total current density. $\nabla$ is the nabla operator. The quantity $\nabla\mathbf{S}$ is nothing but the divergence of the Poynting vector. In a finite volume, in vacuum, where there is no current density, the variation of energy is equal to the flux of the Poynting vector on the surface limiting the volume. There is thus a direct link between linear momentum and energy. In optics, since free electrons can't respond at the optical frequency, there is no exchange between light and the current density. Thus, the same direct relation also holds. The transfer of linear momentum leads to a force. Its action is usually measured by a movement of matter (see \ref{linear}). 

The last quantity to be introduced in this section is the angular momentum of light. It is a vector quantity that expresses the amount of dynamical rotation present in the electromagnetic field of the light. The density of the total momentum of light $\mathbf{J}$ is expressed as follows  \cite{Enk2004}
\begin{equation}\label{eq5}
\mathbf{J}=\epsilon_0\mathbf{r}\otimes(\mathbf{E}\otimes\mathbf{B}) ,
\end{equation}
where $\mathbf{r}$ is the distance between the point where we evaluate the angular momentum density and the origin of the coordinate axis. $\mathbf{J}$ is nothing but the vector product of the Poynting vector with $\mathbf{r}$. This is also a conserved quantity. In most cases, it can be separated into two terms, one corresponding to the Spin Angular Momentum (SAM) associated with the circular polarization of light and the other associated with the Orbital Angular Momentum (OAM) of light. Actually, the problem of separating OAM and SAM is still a debated issue. It has been discussed by several authors \cite{Enk2004,Iwo2011,Klimov2012,Cameron2012}. Nevertheless, within the paraxial approximation, this separation always makes sense. The purpose of this review is not to discuss this separation and we will assume that we can always treat beams within the paraxial approximation. Then the SAM $\mathbf{S}$ density can be written
\begin{equation}\label{eq6} 
\mathbf{S}= \epsilon_0  \mathbf{E_{\bot}} \otimes \mathbf{A_{\bot}},
\end{equation}
where $\mathbf{E_{\bot}}$ and $\mathbf{A_{\bot}}$ are the transverse components of the electric field and the potential vector, respectively. The OAM density $\mathbf{L}$ is
\begin{equation}\label{eq6} 
\mathbf{L}= \epsilon_0 \sum_{i=x,y,z}E^i (\mathbf{r} \otimes\nabla) A^i,
\end{equation}
the $i$-superscripted symbols denote the Cartesian components of the corresponding vectors. It has to be noted that the sum of the OAM density and the SAM density equals the angular momentum density ($\mathbf{L}+\mathbf{S}=\mathbf{J}$). Thus OAM can be deduced from angular momentum and SAM. Note also that there is no relation such as equation \ref{eq4} between the angular momentum density and the energy density, or the linear momentum. As will be seen in section \ref{am}, its detection is performed via rotation of an object. 

\section{Energy measurements}
\label{energy} 

The energy detection corresponds to the integration of the density of energy $u$ introduced in the preceding section (section \ref{theory}, equation \ref{eq1}) over a finite volume. The measured power is the density of energy passing through a finite surface per unit time. 
  
Most of the time, light is characterized via an energy or a power detection. It could be either with the naked eye, or with a photodiode, or with a photomultiplier. It corresponds to the conversion of electromagnetic energy into another kind of energy (internal energy such as atomic or molecular transition, chemical energy, ... ). This energy can then be transformed into a current. Other systems such as bolometers directly transform light energy into a thermal energy that can be also then converted into a current. Note that this current results from a transformation of an energy that has not a direct electromagnetic origin, into a current. It is very different from the current density introduced in equation \ref{eq4}. In this section we will briefly discuss the different systems used to evaluate the energy (or the power) of light. 

\subsection{Bolometer}   
\label{bolometer}
A bolometer is a device for measuring the power of incident electromagnetic radiation via the heating of an absorbing material. This heating can be evaluated with temperature-dependent electrical resistance. The measure only depends on the light power. It is independent of the electromagnetic wavelength. It was invented in 1878 by the American astronomer Samuel Pierpont Langley \cite{Langley1881,Barr1963,Rogalski2012}. Nowadays, it is the most accurate and absolute characterization of small light powers (from the nanoWatt to the milliWatt). 

\begin{figure}
\centering
\includegraphics*[width=8cm]{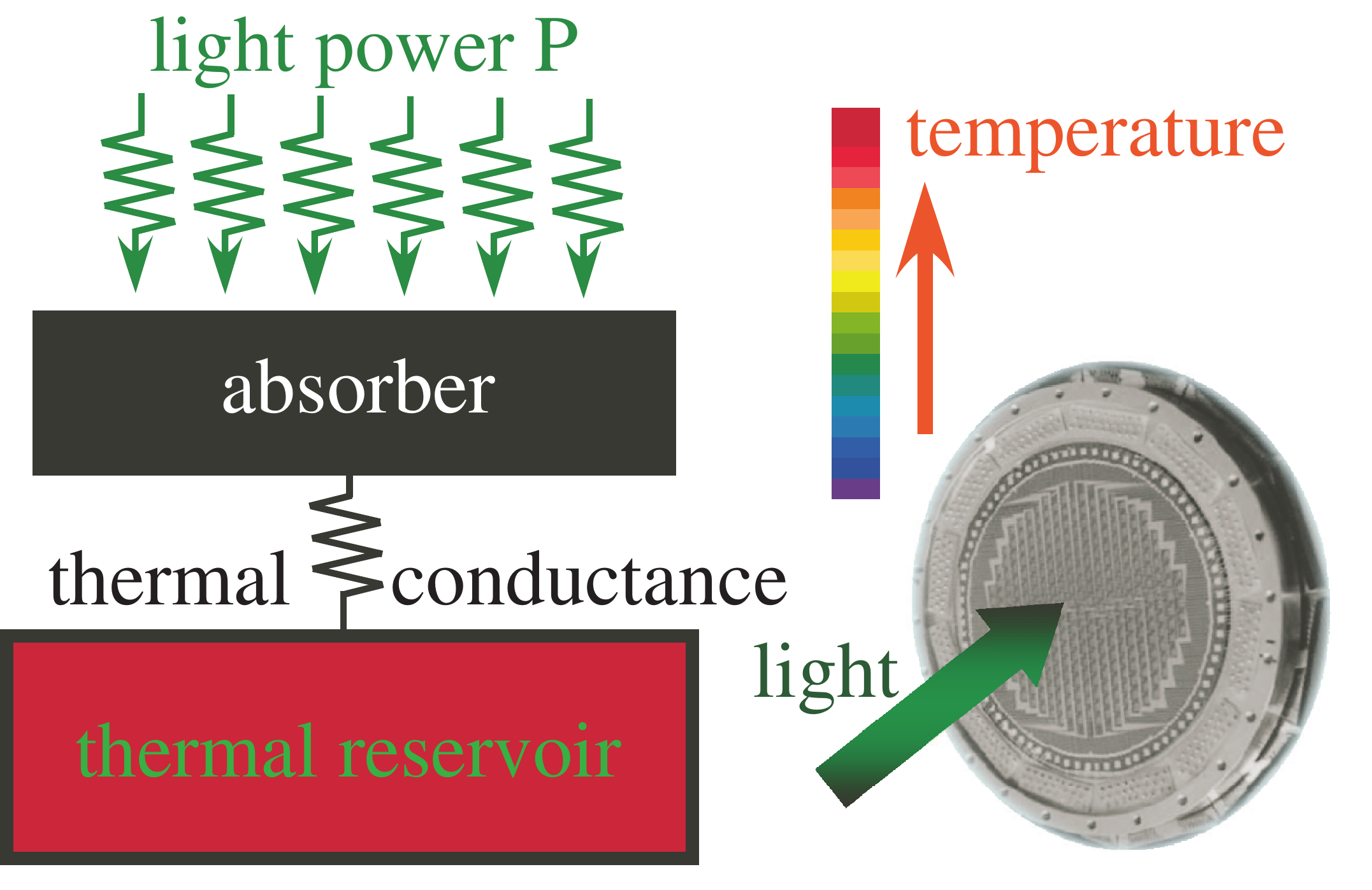}
\caption{Conceptual sketch of a bolometer: the radiated light power P impinges on an absorber connected to a thermal reservoir via a thermal conductance. The current is directly related to the incident light power. Right: example of a typical bolometer that can then be connected to a galvanometer for example.}
\label{fig1}
\end{figure}

In more details, a bolometer consists of an absorptive element (see figure \ref{fig1}), such as a thin layer of metal, connected to a thermal reservoir (at a constant temperature) through a thermal link. Any radiation impinging on the absorptive element raises its temperature above that of the reservoir. The temperature modification can be measured directly with an attached resistive thermometer, or by a thermocouple, or even by the resistance of the absorptive element itself that can be used as a thermometer \cite{Richards1994}. 

 Bolometers are thus directly sensitive to the energy left inside the absorber. Accurate bolometers are very slow to return to thermal equilibrium with the environment. On the other hand, they are extremely efficient in energy resolution and in sensitivity. Note that bolometers are sensitive to any kind of radiation, even to electromagnetic waves in the radio domain. Since they are sensitive to energy, they can also be used in single particle detection, such as $\alpha$-particles or ions \cite{Buhler1988}.

\subsection{Photomultipliers} 
\label{PM}  
The principle of photomultipliers is based on the photoelectric effect. Its first demonstration  was performed by Hertz in 1887 using ultraviolet light \cite{Hertz1887}. Elster and Geitel  demonstrated the same effect using visible light \cite{Elster1889}, two years later. However, historically, the photoelectric effect is associated with Albert Einstein. He advanced the hypothesis that light propagates in discrete wave packets (photons) to explain experimental data of the photoelectric effect \cite{Einstein1905}. He received the Nobel Prize in 1921 for this explanation. Note however that there is no need to invoke the quantization of light and photons to explain the photoelectric effect \cite{Lamb1969}. It can be explained using a classical picture of the electromagnetic field.

The photoelectric effect consists of the emission of electrons when light is shined on a material. However, electrons are emitted only if the frequency of light reaches or exceeds a threshold. Below this threshold, no electrons are emitted from the material, regardless of the light intensity or the time of exposure. Electrons emitted in this way are called photo-electrons. They can be multiplied by a number of electrodes called dynodes in a vacuum tube, up to the anode. This then leads to an output current \cite{Sommer1980}. This current is proportional to the input power, as for the bolometer. The rise time of such photomultiplier tubes can be as fast as several nanoseconds. 

Photomultipliers are associated with the detection of weak light signal and can be operated in single event (photon) counting mode. They are used, for example, in various medical devices to determine the relative concentration of components in blood analysis, for example \cite{Watase2013}.

\subsection{Photodiodes} 
\label{SC}  
\begin{figure}
\centering
\includegraphics*[width=8cm]{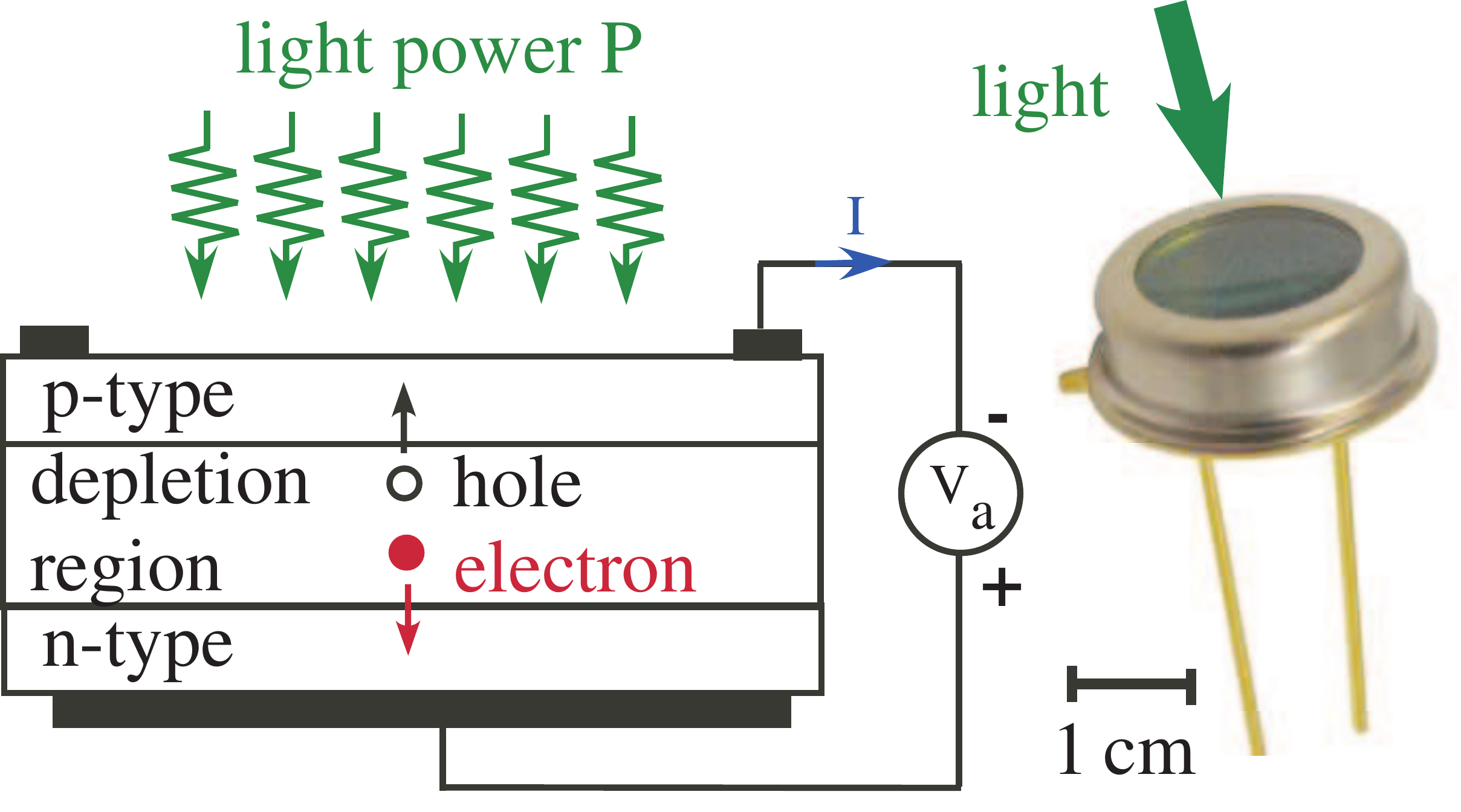}
\caption{Principle of a PIN photodiode. The depletion region is in between p- and n- type doped substrates connected respectively to the anode and the cathode. Right: example of a packed photodiode, showing its typical dimensions. Note that it is the packaging that consumes most of the space.}
\label{fig2}
\end{figure}

A photodiode is a semiconductor device that converts light into current. This current is generated when light is absorbed by the photodiode. More precisely, a photodiode is a p-n junction \cite{Cox2001,Knoll2010,Tavernier2011}. When light with enough energy (typically above the band gap of the semiconductor) hits the diode, an electron-hole pair is created. If the absorption occurs in the junction's depletion region (the region between the n and the p junction), or within one diffusion length away from it, the carriers (electron-hole) are removed from the junction by the electric field of the depletion region. Holes travel towards the anode, and electrons travel towards the cathode (see figure \ref{fig2}). This produces a photocurrent that can be detected or amplified. This photocurrent is proportional to the incident power as for the preceding devices. Depending on the semiconductor used, photodiodes cover the near ultraviolet to the mid-infrared wavelength ranges. 

Photodiodes are used in everyday life. For example, photodiodes govern the closing or opening of automatic doors, they are also used as presence detectors in room lighting. They are much cheaper and easier to use than the photomultipliers. They can reach the same sensitivity with nearly the same rise time. Their dimensions are much smaller. As for the photomultiplier, the response of the photodiode depends on the wavelength. Both have also to be calibrated against a bolometer. 

Photomultipliers tubes and photodiodes also aim at detecting very small fluxes of light down to the single photon detection limit. The development of silicon oxide semiconductor structures with avalanche breakdown operation (also called avalanche photodiodes) has led to single photon visible light detection. The implementation of metal resistive semiconductor structures instead of oxide layers enables the recharge of the structure after photon detection \cite{Zappa1996,Sadygov2000}. This gives high and stable amplification for single photon detection. Silicon photoelectron multipliers are much more sensitive than photodiodes \cite{Saveliev2010}. This new generation of single photon detection finds applications for example, in the wave/particle nature of light \cite{Jacques2007}. The next challenge for single photon detection would be to improve their efficiency, bringing it closer to the quantum efficiency value. 

\subsection{Human eye} 
\label{eye}  

This subsection deals with the human eye, however, the eye of most of the vertebrates responds the same way under the same mechanisms. More generally, vision is based on the absorption of the energy of the electromagnetic field by the photoreceptor cells in the eye. These cells are sensitive to a narrow region of the electromagnetic spectrum, corresponding to wavelengths between 400 and 800 nm. Humans, as most of the vertebrates, have two kinds of photoreceptor cells. These are called rods or cones because of their specific distinctive shapes. Cones function in bright light and are linked with color vision. Rods respond in dim light and are not sensitive to color \cite{Goldstein2001}. 

Let us look a little closer on the mechanisms of light detection in the rods. Rods are narrow elongated elements. The most outer part is responsible for photo-reception. From a chemical point of view, they contain a stack of several (about a thousand) disks. They are wrapped in membranes and packed together with photoreceptor molecules. These photoreceptor molecules in rods are rhodopsin. It consists of one opsin protein linked to 11-cis-retinal, a prosthetic group \cite{Berg2002}. Wald and his coworkers showed that light absorption results in the isomerization of this 11-cis-retinal group of rhodopsin to its all-trans form \cite{Wald1967,Wald1968}. The cis to trans modification of the rhodopsin conformation causes one base nitrogen atom to move from approximately 5 \AA. In essence, the light energy of a photon is converted into atomic motion. The change in atomic positions, sets in train a series of events that lead to the closing of ion channels and the generation of a nerve impulse. This nerve impulse is then transmitted to the brain. 

Cone cells, like rod cells, contain visual pigments. Like rhodopsin, these photoreceptors utilize 11-cis-retinal as their chromophore. The basic principle is exactly the same as for rods cells. The maximum absorption depends on the chemical structure \cite{Berg2002}. In human cone cells, there are three distinct photoreceptors with absorption maxima at 437, 533, and 564 nm, respectively. These absorbances correspond to the violet, green, and yellow regions of the spectrum. They define the blue, green and red perception sensation, respectively. They also correspond to the transfer of the energy of light into an electrical signal. As for other energy detectors, the detection of light by the eye leads to the transformation of the energy into an electrical current (nerve impulse) that is different from the current density that appears in Maxwell's equations. The eye is the most developed human sense.

\section{Linear momentum measurements}
\label{linear} 
In the preceding section (section \ref{energy}), we have reviewed some of the systems or apparatuses used to detect electromagnetic fields, based on an energy observable. Curiously, in the radio domain, most of the detection is performed via the Poynting vector. For example, within an antenna, the Poynting vector excites the electrons that oscillate at the electromagnetic frequency (see equation \ref{eq4}). The energy of the radiated part is equal to the electrical energy. Unfortunately, at optical frequencies the free electrons can't respond anymore. The detection can't be performed via linear momentum transfer to direct electrical current. Most of the electromagnetic field detection in the optical domain is performed via energy measurements and hardly ever through linear momentum detection. 

\subsection{Nichols radiometer}
\label{Nichols} 
Nevertheless, there are other manifestations of the linear momentum (which is proportional to the Poynting vector in the case of plane waves) such as the radiation pressure of light. This is the pressure exerted upon any surface exposed to electromagnetic radiation.  Kepler was one of the first to put forward the concept of radiation pressure in 1619 \cite{Kepler1619}, to try to explain the observation that a tail of a comet always points away from the Sun. The prediction that light has the property of a linear momentum and thus may exert a pressure upon any surface it is exposed to, was made by Maxwell in 1862 \cite{Lewi1908,Page1918}. It has been experimentally proven by Lebedev in 1900 \cite{Lebedev1901} and independently by Nichols and Hull in 1901 with a much better precision \cite{Nichols1901,Nichols1903}. 

\begin{figure}
\centering
\includegraphics[width=8cm]{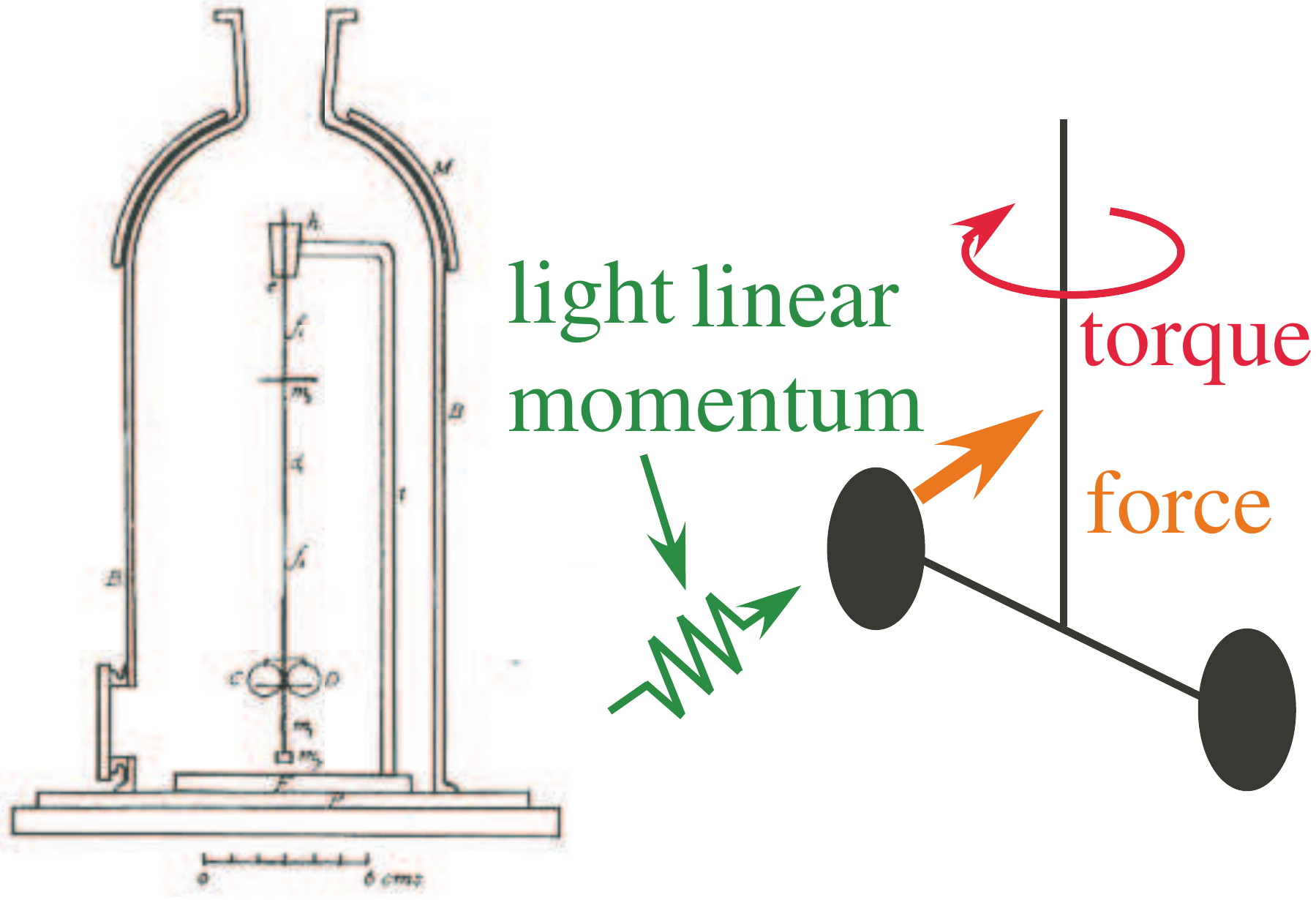}
\caption{Left: picture of the Nichols radiometer (from\cite{Nichols1901,Nichols1903}). Right: principle of a Nichols radiometer, zooming the pendulum inside the vacuum. The light impinging on the vane transfers its linear momentum to the rotating frame, leading to a force and a torque.}
\label{fig3}
\end{figure}

This radiation pressure and the force are very feeble. The force is in the nanoNewton range for a 1 Watt input. Nevertheless, it can be detected as it falls upon an absorbing or reflective metal structure (see figure \ref{fig3}) that can convert the force into rotation. To prevent from any damping from the air, the whole system has to be placed under vacuum. Note that the vacuum within the experimental apparatus has to be quite good otherwise thermal effects may be responsible for a signal detection as in the Crookes radiometer \cite{Crookes1876,Woodruff1966,Guemez2009}. 

Sometimes, the Crookes radiometer is presented as a clear evidence of the manifestation of the radiation pressure of light. This is not correct. The light absorption leads to a local heating and then to a higher pressure (in the thermodynamic sense, the residual pressure being of the order of 1 Pa) on one side of the metal compared to the other. The system then starts to rotate. It is a thermal effect on the radiometer, not an effect due to the linear momentum of light. 

Although it is not often used nowadays, the Nichols radiometer is one of the building blocks of modern opto-mechanical studies. It is clearly recognized as the starting point for nearly all modern radiative force techniques in the manipulation of atoms, particles and macroscopic bodies that will be evoked in the next subsection (subsection \ref{solar}).

\subsection{Other manifestations of the radiation pressure}
\label{solar} 
There are several applications of radiation pressure. One of them is the slowing and cooling of atoms using laser light \cite{Ashkin1978,Chu1998,Cohen1998,Phillips1998}. As a moving atom absorbs a resonant light (corresponding to a transition between atomic levels) with a linear momentum opposed to its own velocity, it is slowed down. When this mechanism operates in three dimensions, atoms can be cooled. Low temperatures below the millikelvin range have been then obtained within a magneto-optical trap or in optical molasses, where the atoms are stuck in light fields and move with difficulty like a spoon in a pot of molasses. 

Radiation pressure is also at the basis of particle trapping. Indeed, particles can be trapped at the focus of tightly focused Gaussian beam where the electric field is maximum. These traps are usually called optical tweezers \cite{Ashkin1986,Grier2003,Mameren2018}. Any dielectric particle experiences a force that moves it towards the beam focus. These particles are indeed high field seekers. Combined with additional forces originating from light scattering and gravity, the resulting force provides a stable trap position for the particles in the vicinity of the focal point of the light beam. They can then be manipulated or trapped with applications in biology and medicine, for example \cite{Fazal2011}.

Radiation pressure has also been used to bend liquid interfaces, although one has to compensate for the surface tension. This can be performed either with high power lasers \cite{Ashkin1980}, or with two liquids with similar surface tensions \cite{Delville2001}, or using parametric amplification \cite{Emilea2014,Emilea2016}, or even via total internal reflection \cite{Emile2011}. In these cases, experimentally, the force is always oriented from the higher index medium towards the lower index medium, as can be seen from the deformation of the interface. 

Cold atoms have applications in several domains in todays life. For example the Global Positioning System (GPS) used in mobile phones relies on time synchronization obtained from atomic clock using cold atoms with very high sensitivity. Research on optical tweezers is now an exponentially growing field with commercially available tweezers. The manipulation of interfaces with light has paved the way to a new exponentially growing domain of physics called optofluidics \cite{Psaltis2006}. The use of light enables new functions in microfluidic devices. 

It has also been proposed to be used in solar sails, following the ideas of Jules Verne in his 1865 book "From the Earth to the Moon" \cite{Tsuda2011,Les2012}. There have been recently several attempts to measure radiation pressure forces. For example, it has been proposed to focus a laser beam at the end of a cantilever \cite{Ma2015} and to modulate the radiation pressure force to separate it from photothermal effects. However, one has to evaluate the spring constant of the system and the cantilever's absorptivity and reflectivity. Nevertheless the system can be used to efficiently measure tiny optical forces using very sensitive devices, i.e., in the picoNewton range. 

Other proposals use dust particles as in an electromagnetic balance, to estimate the light induced force on these particles \cite{Abbas2003}. Indeed, the extra radiation pressure force is balanced by an electrostatic force. This precisely measures the light force on the dust particles which is of great interest in astrophysical studies, for example. However, this gives little information on the characteristics of the electromagnetic field itself. The measurement of radiation pressure forces is also of great importance in delicate equipments using high power lasers to test fundamental phenomena such as gravitational wave detection \cite{Chickarmane1998,Hirose2010}. This measurement could be performed using a Fabry-Perot cavity \cite{Corbitt2006}, looking for instabilities inside the cavity, with a high sensitivity.

Apart from the Nichols radiometer or from the more sophisticated cantilever measurements, all these applications aim to estimate specific consequences of, or give mostly qualitative indications on, the radiation pressure. In particular, they demonstrate its reality, but, hardly give quantitative information on the Poynting vector of the electromagnetic field itself.

\subsection{Abraham-Minkowski controversy}
\label{abraham} 
\begin{figure}
\centering
\includegraphics[width=8cm]{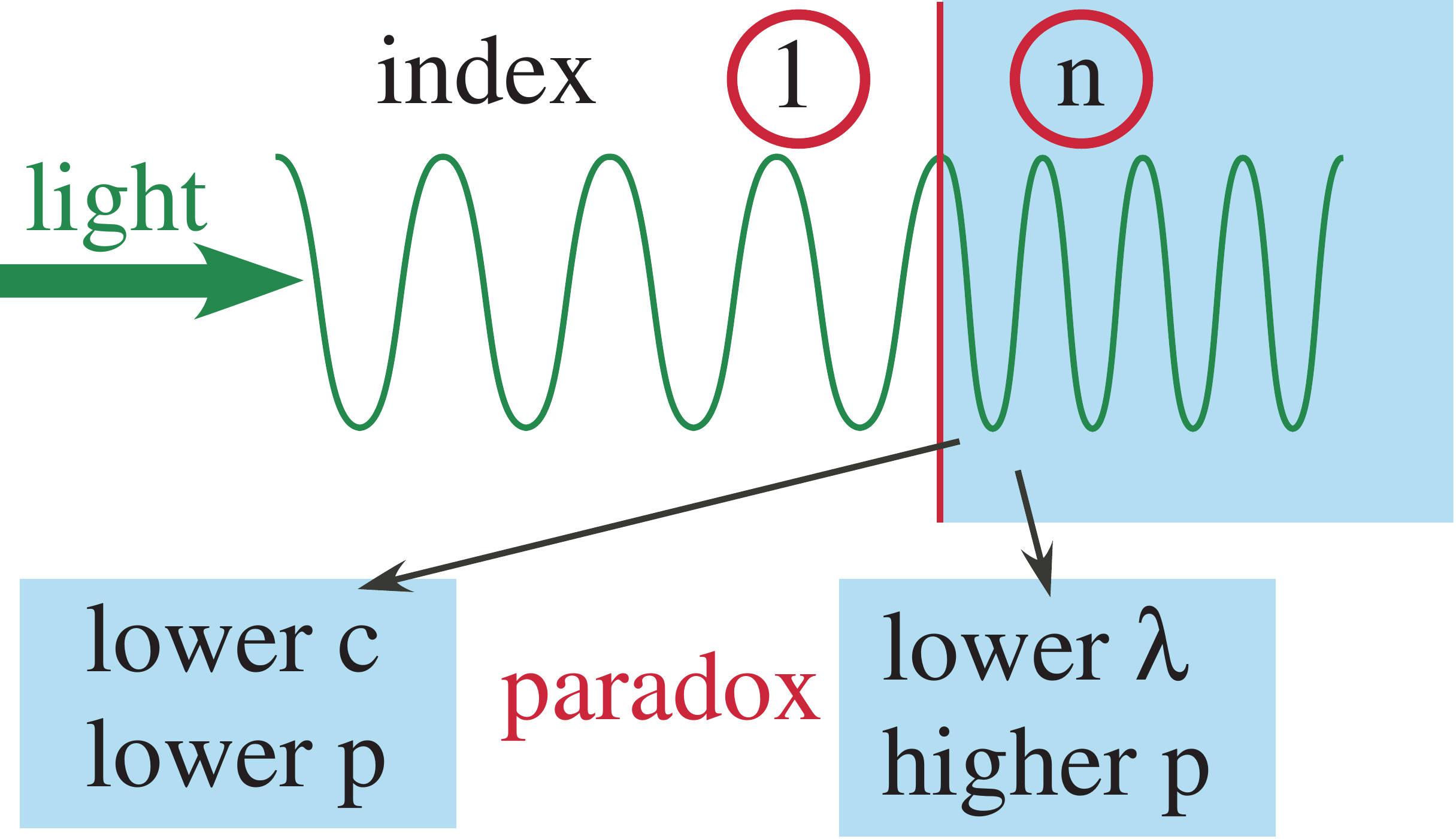}
\caption{At the interface between two media with different optical indexes (for example 1 and $n$, $n>1$), considerations on the change of velocity of light ($c$) or change of wavelength ($\lambda$) lead to a decrease or an increase of the linear momentum ($p$) of light, respectively. Paradoxically, both conclusions seem correct, leading to the so-called Abraham-Minkovski controversy.}
\label{figam}
\end{figure}

The Nichols radiometer is hardly ever used nowadays. The other applications presented in the previous subsection (subsection \ref{solar}) use the radiation pressure as a tool. However, in order to characterize the electromagnetic fields, energy measurements are preferred. There is no need to know it precisely. Nevertheless, quantitative measurements of the linear momentum have been performed recently within the framework of the Abraham-Minkowski controversy. This controversy is rooted in the theory of electromagnetism in matter. It is a fundamental problem about the linear momentum of light. It deals with the way to describe linear momentum transfer between electromagnetic field and matter. The reader can refer to \cite{Gordon1973,Loudon2002,Griffiths2012,Brevik2017} for a review. This debate has been characterized by Ginzburg as a "perpetual problem" \cite{Ginzburg1970}. Briefly, at the interface between two media with different indexes (see figure \ref{figam}), on the one hand, one can consider that in the higher index medium the velocity of light is lowered, leading to a lower linear momentum. On the other hand, one can also consider that the wavelength is lowered leading to an increase of the linear momentum. Both points of view seem correct, but, they are mutually incompatible leading to a paradox. 

In a more formal way, in 1903, Abraham noted that the Poynting vector in matter is \cite{Abraham1903,Abraham1909}
\begin{equation}\label{eq8} 
\mathbf{S}=\frac{c}{4\pi}\mathbf{E}\otimes\mathbf{H},
\end{equation} 	
where $\mathbf{B}=\mu\mathbf{H}$, ($\mu$ is the magnetic permeability in the medium, an $\mathbf{H}$ is the magnetizing field), following equation \ref{eq3}, the density of linear momentum is
\begin{equation}\label{eq9} 
\mathbf{p}=\frac{1}{4\pi c}\mathbf{E}\otimes\mathbf{H},
\end{equation} 	
whereas, in 1908  Minkowski gave the following  alternative derivation of the electromagnetic momentum density \cite{Minkowski1908}
\begin{equation}\label{eq10} 
\mathbf{p}=\frac{1}{4\pi c}\mathbf{D}\otimes\mathbf{B},
\end{equation} 
where where $\mathbf{E}=\epsilon\mathbf{D}$, ($\epsilon$ being the electric susceptibility in the medium, and $\mathbf{D}$ is the displacement field). 

On the one hand, the Minkowski formulation is usually associated with the recoil momentum of absorbing or emitting guest atoms in a host dielectric and represents the combination of both field and material momentum values. It is sometimes called the canonical momentum. On the other hand, the Abraham momentum is associated with the kinetic momentum and represents the photon momentum without any material contributions \cite{Nelson1991,Milonni2010,Sheppard2016}. Considering plane waves in an homogeneous medium of a given index $n$, Abraham photon momentum is inversely proportional to $n$, while Minkowski photon momentum is directly proportional to $n$ (see figure \ref{figam}). The two visions seem not to be compatible. 

There have been several measurements of the linear momentum density \cite{Walker1975,Campbell2005,She2008,Wang2011,Choi2017}, supporting either Abraham's or Minkowski's vision with qualitative or quantitative agreement. There has been even one recent report \cite{Zhang2015} showing that both Minkowski and Abraham pressure of light have been observed on the same experiment.   

Finally, it turns out that, apparently, both forms are correct, but represent different types of momenta \cite{Milonni2010,Sheppard2016,Pfeifer2007,Mansuripur2009,Mansuripur2010,Barnett2010,Kemp2011,Mansuripur2013}. Both of them can be measured. It depends on the definition of the system at the heart of linear momentum transfer. The total momenta of matter and light are conserved, but its division into optical and material parts is arbitrary. It may be performed so as to separate kinetic or canonical parts. Both of them are physically meaningful, despite the determination of the kinetic part is sometimes difficult from a statistical physics point of view. Indeed, it is nearly impossible to describe each part of matter individually. It has to be defined from a statistical point of view. Nevertheless, depending on the way the system is defined or considered, they may thus apply under different experimental conditions. 

\section{Angular momentum measurements}
\label{am} 
In the preceding sections (sections \ref{energy} and \ref{linear}), we have discussed the detection of energy and linear momentum. Both observables are linked by equation \ref{eq4} which correlates the energy density with the linear momentum density. Except in the few cases discussed above, energy is the most often quantity measured to characterize the electromagnetic field. 

However, there is another quantity, the angular momentum, that is also a conserved quantity. It is independent from the two other observables. It also partly characterizes the electromagnetic field. It is linked to a rotational aspect of the electromagnetic field. Within the framework of the paraxial approximation, this angular momentum can be divided into SAM that is linked to the circular polarization of light (either left or right), and the OAM of light that characterizes the rotation of the Poynting vector along the direction of propagation. 

\subsection{Spin angular momentum}
\label{sam}
SAM is linked to the light polarization that is known probably since the Vikings \cite{Ropars2012}, and surely from the $19^{th}$ century \cite{Malus1810}. It has several applications ranging from communication to polarization microscopy, and more recently cinema 3D technology. The polarization of light offers many applications in the daily life. Although OAM has gained considerable interest in the recent years, with application in various domains such as optical micromanipulation, quantum optics, communications, and radar, people are usually more familiar with SAM. It is also of more common popular use than OAM.  

Indeed the SAM is associated with the circularly polarized light that can be either right or left polarized (or equivalently, $\sigma+$ or $\sigma-$, or clockwise and counterclockwise). When the light beam is linearly polarized, there is no SAM. Most of the time (except for the rotations of particles or objects described below) the detection of the circularly polarized light is performed with a quarter wave plate ($\lambda/4$) that transforms a circularly polarized light into a linearly polarized light. This linearly polarized light is then detected via a linear polarizer and a detector dedicated to the optical power detection. This can be schematized by a filter (polarizer) and a detector of energy like the ones described in section \ref{energy}. The same mechanism also holds for animals that use polarized light in vision such as bees \cite{Frisch2014}, octopuses \cite{Shashar1996}, or other animals \cite{Horvath2013}. One can also note the recent development of a detector directly sensitive to polarized light \cite{Li2015}. It is a single ultra compact element that uses chiral plasmonic metamaterials (i. e. a material engineered so as to have properties usually not found in nature) to discriminate between right and left circularly polarized light.  

Polarization plays an important role in light/matter interaction. Matter, such as chiral material or molecules may have a different response to polarized light  \cite{Born1965}. For example, in the case of circular birefringence \cite{Arago}, the index for the right and left polarization may be different, leading to a rotation of a linearly polarized light (also known as optical activity). Similarly, circular dichroism is the differential absorption of left- and right polarized light \cite{Atkins2005}. This may then be used, for example, in circular dichroism spectroscopy \cite{Greenfield2006}. Circular polarized light may be important in magnetic recording \cite{Stanciu2007}.

Moreover, polarized light is at the basis of optical pumping \cite{Kastler1950,Kastler1957}. Light is used, for example, to pump bounded electrons of atoms or molecules into a well-defined quantum state (such as a single hyperfine sub-level). Then the system is said to be oriented. It may then be used as magnetometer \cite{Mhaskar2012}. It may also sometimes generate sharp resonances \cite{Gilles2001}, or lead to applications in light induced transparency \cite{Harris1997}. Nevertheless, in all these examples, the detection of the SAM is performed via energy measurements. 

The only way to directly observe the SAM of light is to detect its mechanical action on a system, i.e. a torque effect due to the transfer of angular momentum from light to matter. The first experimental demonstration has been performed in 1936 by Beth \cite{Beth1936}. It was inspired from the Einstein and de Haas experiment on electrons \cite{Einstein1915}. He demonstrated the transfer of angular momentum from a circularly polarized light to a suspended birefringent plate. There have been several qualitative results using radio frequency radiation  \cite{Carrara1949,Allen1966} where the torque may be higher than in optics. Indeed, people have been able to clearly observe the rotation of a suspended mobile, but establishing a quantitative relation between the measured rotation, the expected torque, and the electromagnetic field is much more difficult. 

Experimental results have also been reported for particles in suspension in liquid where the steady state rotation only, is observed \cite{Simpson1997,Friese1998,Higurashi1998,Freise2001,Patterson2001,Garces2003}. In these specific cases, although presenting clear evidence of the existence of SAM, it is then difficult to access the SAM density itself. In particular, one has to consider the wetting characteristics of the particle-liquid system as well as the flow properties in order to evaluate the friction coefficient that has to be known exactly. This is usually tricky, as we have recently shown for OAM,  \cite{Emileb2016,Emile2018}. For example, the value of the drag coefficient is generally extracted from much data, but depends strongly on the parameters of the model used. Although the results on the rotation of objects are very convincing, steady state rotation measurements are not very quantitatively accurate.

\begin{figure}
\centering
\includegraphics[width=8cm]{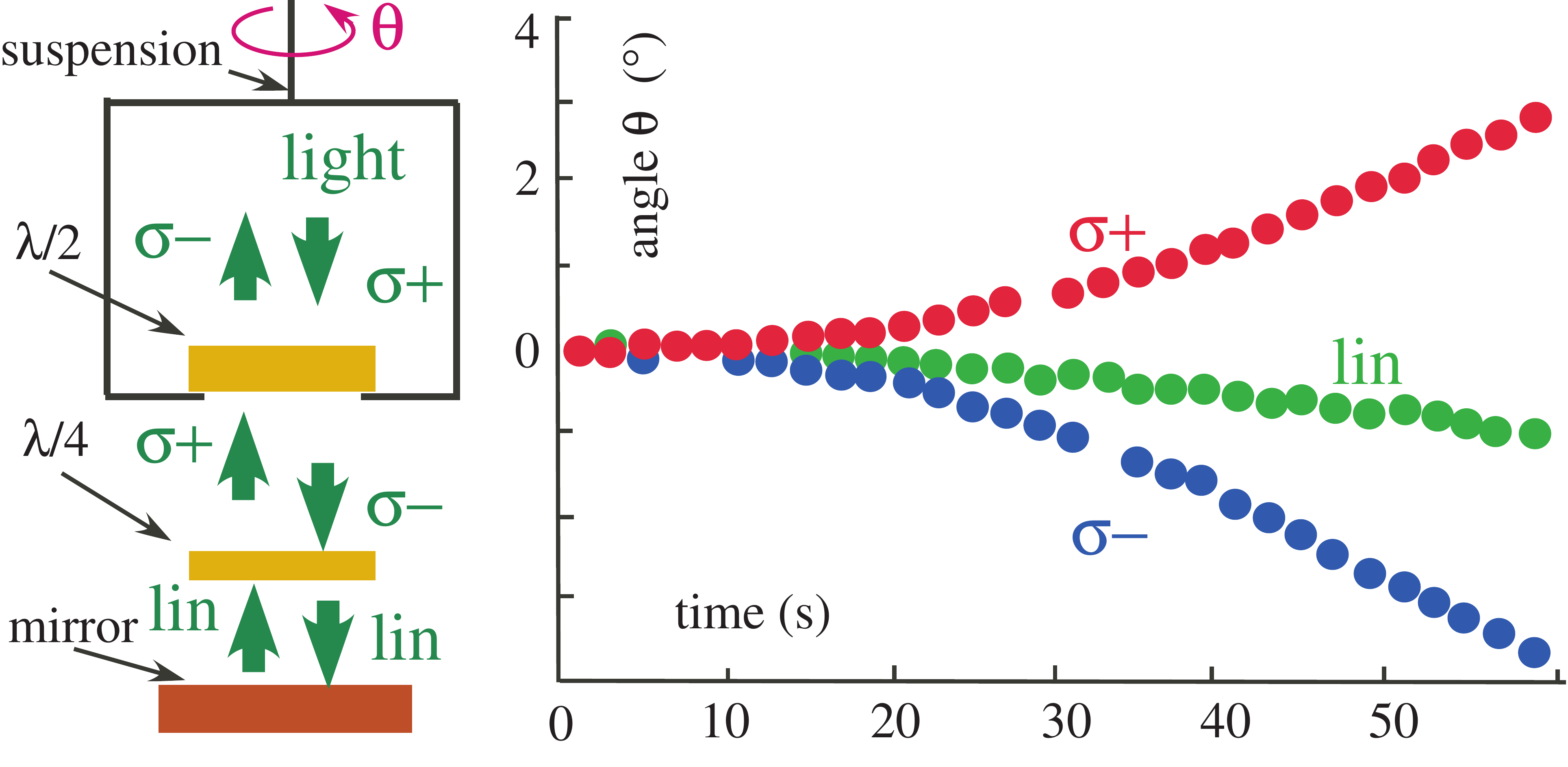}
\caption{Left: Principle of the experiment. It corresponds to a $\sigma+$ incident light. The first $\lambda/2$ changes the polarization to $\sigma-$. Then light crosses twice a $\lambda/4$ that is equivalent to a $\lambda/2$. The light is $\sigma+$ polarized and finally transformed to $\sigma-$ after the $\lambda/2$. The signs are reversed for a $\sigma-$ incident light. Note that such a configuration realizes a so-called helicoidal wave\cite{Siegman1965,Kastler1970}. Right:  observation of the accelerated regime for $\sigma+$, $\sigma-$ and linear polarization of the incident light.}
\label{fig4}
\end{figure}

On the other hand, observations in the uniformly accelerated regime, with a negligible damping (the experiment is performed in air or in vacuum), lead to measurements of the acceleration and then of the torque. This was already done in Beth's experiment \cite{Beth1936}. However, quantitative results are difficult to obtain. Beth used an indirect technic namely parametric amplification and he was only able to observe the sense of rotation. The direct observation of the spin transfer has been reported using a CO$_2$ laser at a wavelength $\lambda=10.8$ $\mu$m \cite{Delannoy2005} (see figure \ref{fig4}).  

Instead of using the light absorption, a birefringent half waveplate $\lambda/2$ is used that reverses the handedness of a given circularly polarized light (like in the Beth experiment). This doubles the transfer efficiency. The light then crosses a quarter waveplate ($\lambda/4$, see figure \ref{fig4}) before being retro-reflected. This again doubles the effect. From the acceleration observed in figure \ref{fig4}, and from the estimated inertial momentum $J$ of the frame and the 10 mm-diameter $\lambda/2$ ($J=1.5\pm0.3 10^{-8}$ kgm$^2$), one can estimate the torque. For right circularly polarized light ($\sigma+$ light), and a power of $P=15$ W, the torque is $\Gamma=3.1 \;10^{-13}$ Nm, and for left circularly polarized light ($\sigma-$ light), the torque is $\Gamma=3.7 \;10^{-13}$ Nm, with a reverse rotation. The two absolute values are nearly equal. This experiment is a direct measurement of the transfer of SAM of light to matter, and thus a measure of the SAM. 

\subsection{Orbital angular momentum}
\label{oam}

\begin{figure}
\centering
\includegraphics[width=8cm]{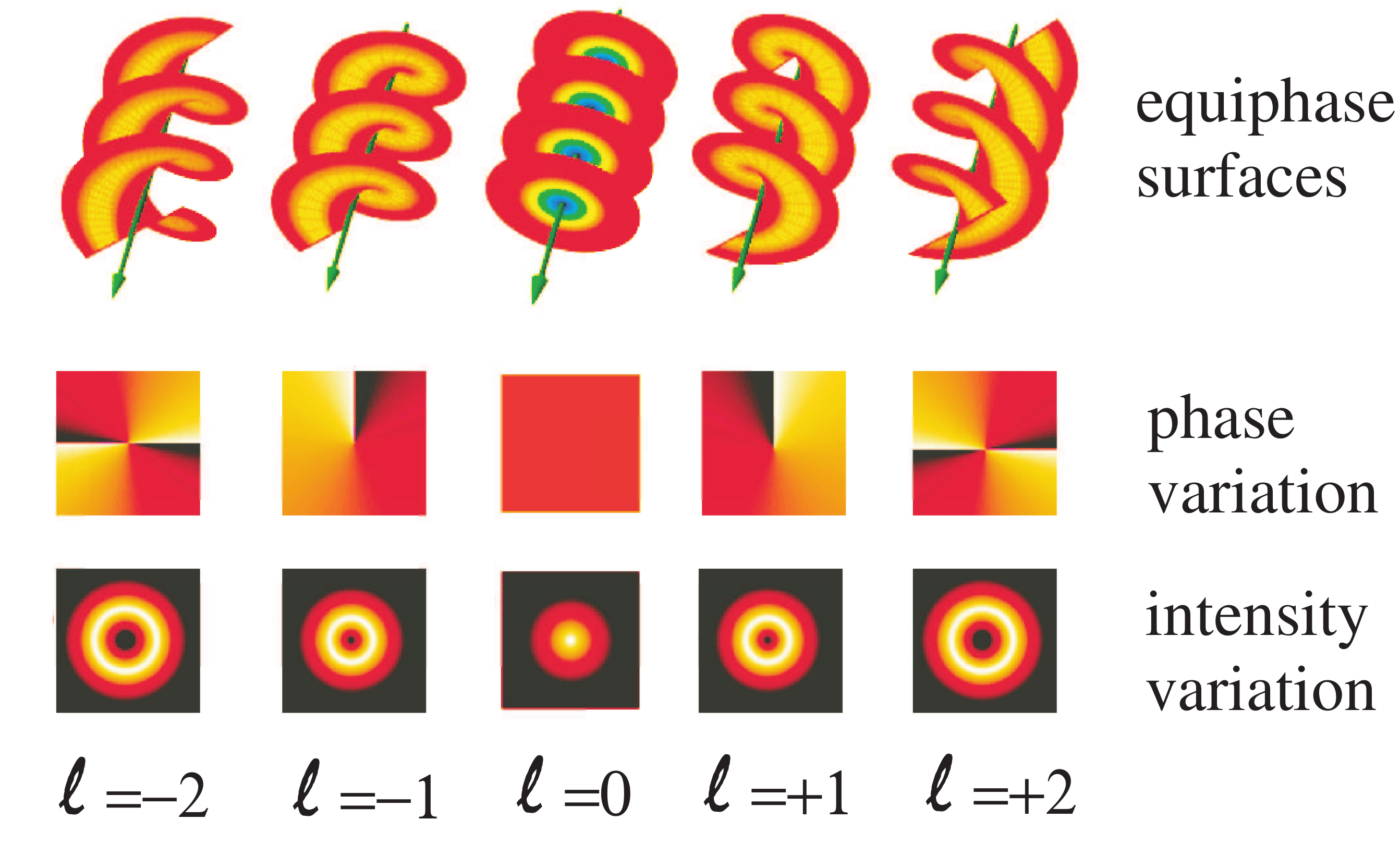}
\caption{Representation of beams carrying OAM: iso-phase surface (top), phase distribution (middle), and intensity distribution (bottom) for $\ell=-2, -1, 0, +1, +2$, in a plane perpendicular to the direction of propagation. It exemplifies the vortex structure. This figure is inspired from \cite{Karimi2009,Padgett2014}.}
\label{fig5}
\end{figure}

The other part of the angular momentum is the OAM. Although it was already described in Poynting's early work \cite{Poynting1909,Jackson1998}, it has gained a great renew of interest in the 90's \cite{Bazhenov1990,Allen1992} and is now a well-established field \cite{Allen2003,Molina2007,Padgett2011,Yao11,Barnett2017}. Usually, an electromagnetic field carrying OAM is described as a beam that has a hole in the center of its amplitude distribution (donut shape), and a phase $\varphi$ that is not uniform (see figure \ref{fig5}). Its phase varies as $\varphi=\ell\theta$, $\theta$ being the polar coordinate and $\ell$ being the so-called topological charge. On a plane perpendicular to the direction of propagation, it has a $2\pi\ell$ variation around the axis of the beam. This beam is also sometimes called a vortex beam, or a twisted beam.

Most of the time, the characterization of such a beam is performed either by transforming the twisted beam into a fundamental Gaussian beam carrying no OAM and thus having a uniform phase, or through interferences. For the former, it can be obtained by operating, for example, the mode creation optics in reverse \cite{Allen1992,Yu1992,Heckenberg1992,Beijersbergen1993,Berkhout2010}, then experimentally demonstrating a uniform phase for the transformed beam.  It can also be achieved via interferences, either with a plane wave \cite{Yu1992,Harris1994,Vickers2008,Lavery2012} or by self interferences \cite{Leach2002,Guo2009,Hickmann2010,Mourka2011,Anderson2012,emile1,emile2,emile5}. 

Nevertheless, theses techniques are able to characterize the phase variation of the twisted beam only. It is not at all related to the fact that the electromagnetic field indeed carries OAM. There is another method that uses the rotational Doppler shift of the beam \cite{Courtial1998,Vasnetsov2003}, that looks for a frequency change of the beam when passing through a rotating medium. This has also been observed for polarization \cite{Garetz1981}. This technique makes partly use of the angular momentum character of the beam. 

Like for the polarization the only way to fully characterize the OAM's rotational nature is by angular momentum transfer via torque measurements. OAM can be transferred to particles that absorb light, making them rotate \cite{He1995}. However, torque measurements are more tricky. As for SAM, measurements from steady state rotation are delicate. Indeed, most of the time the object to be rotated is in suspension in a liquid, or floating at the air/liquid interface. Then, the friction coefficient has to be known or at least eliminated from several measurements \cite{Nieminen2001,Parkin2006} to determine the torque. Alternatively, torque can be deduced from a uniformly accelerated movement, with negligible friction, independently from the power used, with a higher precision, like it has been done for SAM.  
 
\begin{figure}
\centering
\includegraphics[width=8cm]{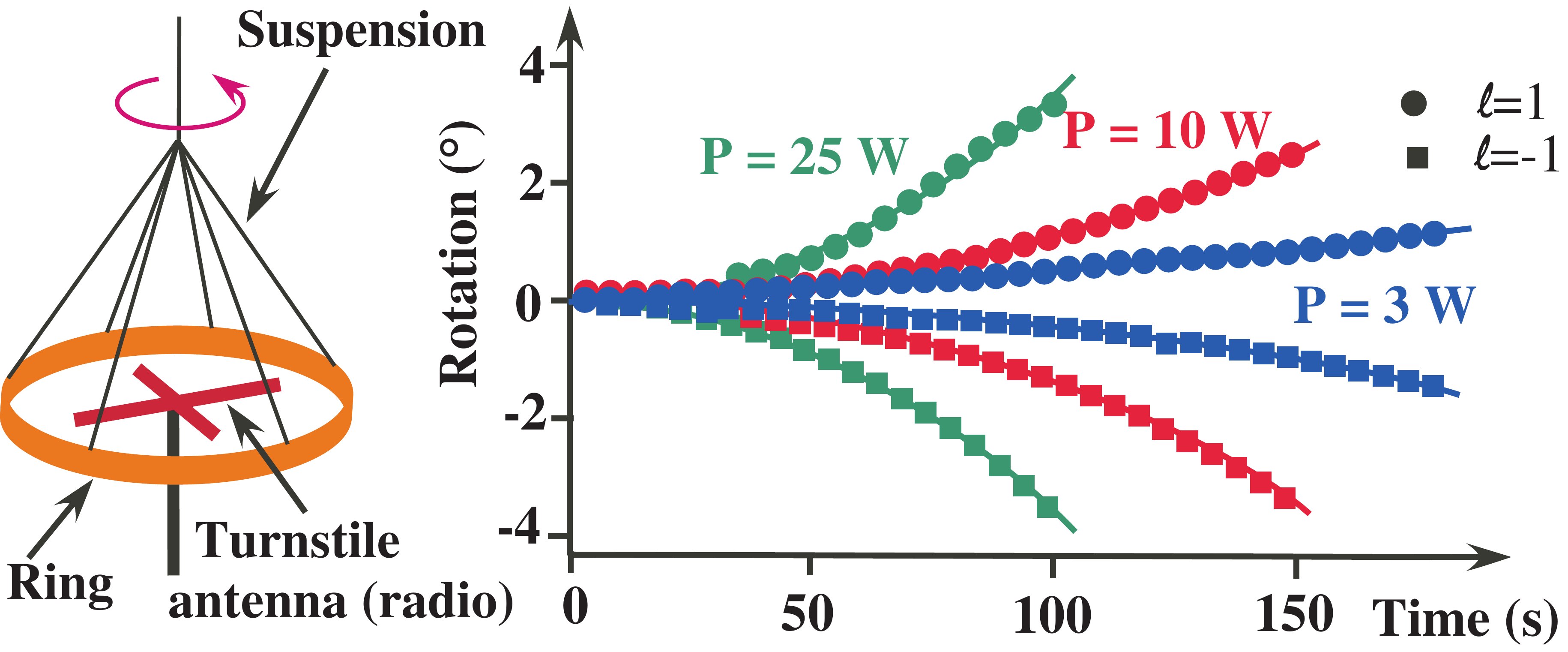}
\caption{Principle of the experiment (left). The radius of the ring is 15 cm. The electromagnetic field radiated by the turnstile antenna is in the radio domain. Accelerated regime for various powers, and for $\ell=-1$ and $+1$ (right).}
\label{fig6}
\end{figure}

This kind of experiment has been realized in radio around a frequency $\nu= 1$ GHz. Actually, since the strength of the effect depends linearly on the wavelength, as for SAM, the torque is higher in radio than in optics  \cite{Emile2014,Emile2016,Emile2017}. A so-called turnstile antenna emits a $\ell=+1$ OAM wave in the plane of the antenna (see figure \ref{fig6}). A suspended ring with a moment of inertia $J=8.4 \; 10^{-4}$ kgm$^2$ reflects the electromagnetic wave. When the electromagnetic field carries OAM, it starts to rotate. The entire experiment is placed in an anechoic chamber. We have been able to demonstrate the transfer of OAM from an electromagnetic wave to a macroscopic object. We observed a uniformly accelerated regime. For example, for a 25 W power, the acceleration equals $7.8 \; 10^{-4}$ $^\circ/s^{2}$ which corresponds to an OAM torque of $ \Gamma=1.1\; 10^{-8}$ Nm.

\begin{figure}
\centering
\includegraphics[width=8cm]{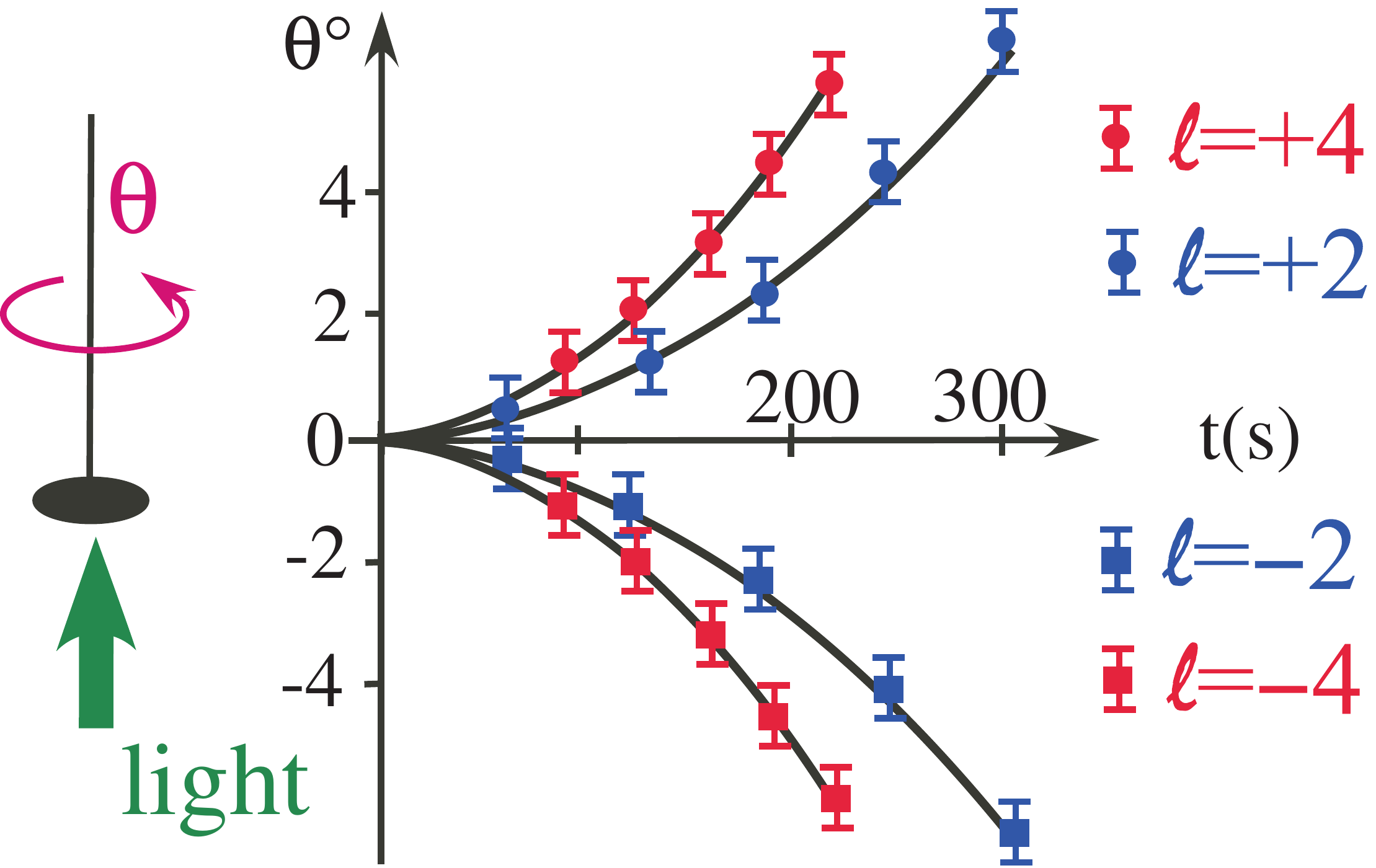}
\caption{Principle of the experiment (left). The diameter of the absorbing object is 1.5 mm. Uniformly accelerated rotation for $\ell=-4, -2, +2$ and $+4$ (right).}
\label{fig7}
\end{figure}

We have also carried out the experiment in optics. It is similar to what has been done for SAM \cite{Delannoy2005}. A 1.5 mm-diameter absorbing black paper (density 180 gm$^{-2}$) hangs from a 10 cm-long ordinary cotton thread. The whole system is set in a vacuum chamber (pressure of 0.5 Pa). The beam to be characterized by the torque measurement is focused on the black paper with a 5 cm focal length ordinary lens. The OAM transfer is here by absorption. We register with a camera (for 6 min at most) the rotation of the suspension (see figure \ref{fig7}) and evaluate the rotation angle. Since it is small and since the thread is long, the restoring torque is negligible. Besides, the system is in a vacuum chamber, leading to a negligible friction. The possible heating of the black paper has no influence on the torque. We observed a uniformly accelerated rotation. We then deduced the angular acceleration $\gamma$. We evaluated the moment of inertia $J$ of the paper. It equals $J=4.473\pm0.003\;10^{-14}$ kgm$^2$. The torque is thus $\Gamma=J\gamma$. 

Independently, we measured the topological charge of the beam using Young's double slit experiment \cite{emile1}. We have carried out the torque measuring experiment for $\ell$ varying from $-8$ to $+8$. Figure \ref{fig7} displays the uniform accelerated regime for $\ell=-4, -2, +2$ and $+4$. For example, for $\ell=+2$, and a measured power $P=0.30\pm0.01$ mW at the paper location (measured before evacuating the chamber), we find a torque $\Gamma=1.00\pm0.05\;10^{-19}$ Nm. The torque is exactly reversed for $\ell=-2$. The measurements using topological charges varying from $\ell=-8$ to $\ell=+8$, lead to the same conclusions. We have also performed experiments using this apparatus for circularly polarized light, i.e. with SAM, and also observed the uniformly accelerated rotation regime. 

Finally, as a conclusion concerning section \ref{am},  the characterization of the angular momentum density (either spin or orbital) can only be performed with a torque mechanical measurement, via its transfer to an object. The precision of the estimation  depends on the observation of the rotation and  of the estimated moment of inertia.  

\section{Discussion and consequences}
\label{discussion}

As already mentioned, in optics, there is equivalence between energy and linear momentum measurements. Indeed, according to equation \ref{eq4}, without current density, energy and linear momentum are linked. It means that in the absence of any current coupled with the electromagnetic field, the energy absorption in a finite volume equals the flux of the Poynting vector through the surface, limiting this volume. Since it is usually easier to measure the energy of light than its linear momentum, linear momentum measurements are hardly ever performed. The ratio of the energy and the linear momentum is equal to $c$ the celerity of light. 

There is no such relation between energy and angular momentum. Indeed, these two quantities are independent, although, in the exchange from light to matter, both energy and angular momentum must be conserved \cite{Kristensen1994,Mansuripur2008}. For example, in the case of Beth's experiment \cite{Beth1936}, exchange of SAM leads to the rotation of a suspended rotating birefringent plate. The plate thus gains energy. One may then wonder where this energy comes from. As already mentioned, energy must be conserved between light and matter, although there is no light absorption. 

Actually, the only way to fulfill the above requirements, since the energy depends on the wavelength, is to consider that the frequency of light is lowered during the exchange. If one considers incident particles of light (i. e. photons), since the number of photons is conserved, their energy must decrease.  This leads to the same conclusion, the light frequency must decrease. This ensures energy conservation \cite{Bretenaker1990}. The conservation of angular momentum must be considered first, before energy conservation. However, the change of energy of light is only an indirect consequence of angular momentum exchange. Because the suspended birefringent plate has gained energy, and because the only source of energy is from the light frequency, the light frequency is lowered. This frequency lowering could indeed also be understood as a rotational Doppler effect \cite{Garetz1981,Courtial1998,Vasnetsov2003}. 

More generally, from a mechanical point of view, as reported by Truesdell \cite{Truesdell1968,Thide2015} on discussions between Euler and Bernoulli, angular momentum is a physical observable in its own right, in general independent of and not derivable from linear momentum or energy. The knowledge of one of these quantities doesn't imply the knowledge of the other (see table \ref{tlab}). Linear momentum and energy on one side and angular momentum on the other side are indeed truly independent quantities. 

\begin{table}
  \centering
  \caption{Conserved quantities, linked between each other and their experimental observation. E: energy, p: linear momentum, L: angular momentum}
  \label{tlab}
  \begin{tabular}{@{}lccc@{}}
    \toprule
    & E & p & L\\ 
    \midrule
    Linked to & p & E & none\\
    Observation & energy transfer & force & torque\\
    \bottomrule
  \end{tabular}
\end{table}

\begin{figure}
\centering
\includegraphics[width=8cm]{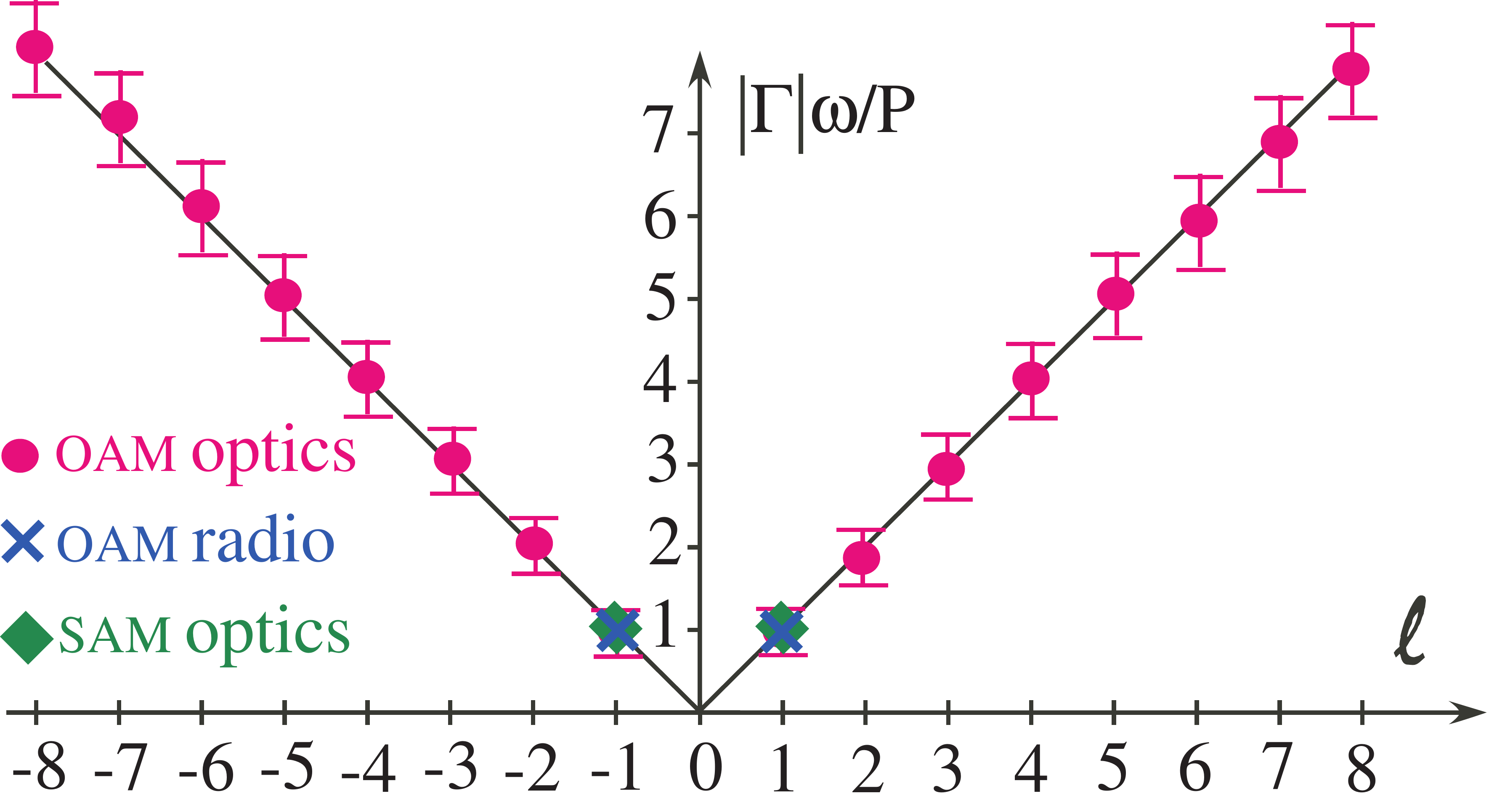}
\caption{The ratio of the product of measured torque $\Gamma$ with the pulsation $\omega$ to the measured optical power $P$,  versus the topological charge $\ell$. Note that we have also included the results of the observations made in radio that superpose with the results for SAM in optics}.
\label{fig8}
\end{figure}

Nevertheless, one must admit that, intuitively, the torque is proportional to the light power. Experimentally, when the power is increased, the torque increases. Could any relationship between the energy and the torque be experimentally found? In particular, let us have a deeper insight on the experimental results of section \ref{am}. We have plotted in figure \ref{fig8}, the ratio of the measured torque times the pulsation of light, to the light power measured at the same place, versus the topological charge (or the circularity of light, $\ell=+1$ for $\sigma+$ light and $\ell=-1$ for $\sigma-$ light). This corresponds to the experimental results of figures \ref{fig4}, \ref{fig6} and \ref{fig7}. Clearly,  
\begin{equation}\label{eq9} 
\frac{\Gamma\omega}{P}=\ell
\end{equation} 	
the ratio of the measured torque times the pulsation of light, to the light power equals the topological charge, which is an integer number. It seems that the topological charge must indeed be quantized. 

For circularly polarized light there is only 2 values for $\ell$, $\ell=+1$ or $\ell=-1$. One can argue that when calculating classically the SAM and the energy, one finds that their ratio is equal to $\omega$ or $-\omega$ depending on the circularity of the polarization. It equals zero for an equally weighted combination of them. However, doing so, one implicitly assumes that the polarization can only be $\sigma+$ or $\sigma-$, or a combination of the two. This reasoning is very similar to the one performed by Raman and Bhagavantam \cite{Raman1931}. This result is linked to the expression of the polarization in the direction of the field propagation. This can be so performed, because spin is a local concept. It is defined locally, at a given position. 

For light carrying OAM, there is no limit to the $\ell$ value in principle, although it is always an integer number. One can also argue that when considering a classical light field with cylindrical symmetry, or equivalently with a given topological charge, one would get the result of equation \ref{eq9}. It seems that there is no need to invoke any quantization of the field. However, considering a cylindrical symmetry or a given topological charge, means that after one turn around the axis of propagation of light, the phase variation must be unchanged, i.e. equal to an integer number (the topological charge) times $2\pi$. People following this reasoning implicitly assume that the OAM is quantized. This kind of reasoning has also been recently used in the case of acoustic waves with radiation pressure and torque \cite{Demore2012}, following theoretical considerations \cite{Hefner1999,Skeldon2008}. To establish that the value of the ratio of the OAM to the radiation pressure, they implicitly assume the existence of the phonon and that acoustic waves are indeed quantized.

Thus, figure \ref{fig8} is an unambiguous signature that the angular momentum is quantized. Since $\hbar$ is the quantum action, one can divide the torque by this quantum of action times the topological charge i.e. $\ell\hbar$. One then finds the number of particles involved in the torque effect per second, each particle carrying an angular momentum equal to $\ell\hbar$. This means that in a light beam carrying OAM, the OAM is transported by individual particles, each of them carrying an integer number of $\hbar$ corresponding to the topological charge. These particles could be assimilated to photons. Indeed from the value of the torque divided by $\ell\hbar$ one deduces the number of particles. This number is also equal to the incident light power divided by $\hbar\omega$, which is usually assumed to be the number of photons in a light beam. Simultaneous measurements of torque and light power lead to a clear experimental demonstration of the quantization of light. 

\section{Conclusions}

Usually, a beam of light is characterized by its power or by its intensity. This is even true for phase measurements performed using interferences, since these interferences are detected via intensity measurements and contrast. Polarization is usually detected with a filter and a power meter or a photodiode. The linear momentum of light can be used to exchange momentum with matter, for example, in laser cooling of atoms or molecules, in trapping particles, in the bending of interfaces, or even in solar sails. It is also used in measurements dedicated to its detection such as in the Abraham-Minkowski controversy. Nevertheless, apart from these examples, the linear momentum is hardly ever used to characterize light beams. The main reason is that energy and linear momentum depend on and can be deduced from each other as can be seen in equation \ref{eq4}.

This is fundamentally different for the angular momentum of light (either SAM or OAM). Angular momentum is independent and can't be deduced from energy and vice versa, although both quantities must be conserved. Nevertheless, the torque generated by light depends on the light power. This can be explained via the number of particles, that can also be called photons, carrying angular momentum. This is also the same as the number of particles carrying energy. Single particles carry $\ell\hbar$ angular momentum and $\hbar\omega$ energy. 

In electromagnetic theory, energy, linear momentum and angular momentum are quantities that are ruled by conservation laws. However, in the electromagnetic theory, there is another quantity that is conserved. It is called the \textit{boost momentum density} \cite{Ribaric1990,Barnett2011,Thide2015}. It is related to the generators of the so-called Lorentz boots in special relativity \cite{Feynman2005,Feynman2005b}. It couples a Cartesian direction with time. In a more formal description, it equals the difference between the moment of energy density and the product of linear momentum density by the elapsed time. For electromagnetic fields in free space, the three components of the boost momentum of energy (one for each of the three orthogonal Cartesian directions) are constant of motion. 

Although people were aware of the existence of angular momentum in light fields they had little interest in it since the beginning of the 90's and the work of Vasnestsov et al. \cite{Bazhenov1990} and Allen et al. \cite{Allen1992}. It is now an exponentially growing field. Up to now, the boost momentum has been hardly ever exploited. Similarly to angular momentum, it may in a near future pave the way to new characterizations and properties of the light with unsuspected and unprecedented consequences in the field of electromagnetism. In particular, as exemplify by Barnett et al. \cite{Cameronb2012} and also more recently by Bliokh \cite{Bliokh2018}, the boost eigenmodes of the boost momentum are related to the Lorentz symmetry. They describe the propagation of relativistic signal. Since they never violate causality, they may thus play an important role in problems involving causality and supraluminic propagation \cite{Sommerfeld1950,Brillouin1960}.      


\begin{acknowledgement}
The authors acknowledge technical support from Jean-Ren\'e THEBAULT from the University of Rennes 1, and useful discussion with Christian BROUSSEAU and Kouroch MAHDJOUBI from the University of Rennes 1. The authors would like to thank Sean MC NAMARA from the University of Rennes 1 for careful reading of the manuscript.
\end{acknowledgement}

\begin{biographies}
\authorbox{cvemileo}{Olivier EMILE}{received his PhD from the University of Paris 6 (France) in 1993, in laser cooling of atoms. He has been appointed as an assistant professor at the University of Rennes 1 in 1994 and as a full professor at the same university in 2002. His main research activity includes laser dynamics, evanescent waves, atomic physics, optofluidics and angular and orbital momentum of light.}
\authorbox{cvemilej}{Janine EMILE}{received her PhD from the University of Rennes 1 (France) in 1993 in condensed matter. She has been appointed as an  assistant professor and as a full professor in 1993 and 2007, respectively, still at the University of Rennes 1. Her research interest focuses mainly in soft matter, colloidal suspensions, optofluidics and liquid films. She has also been interested in studies on the orbital angular momentum of light and its application to particle manipulation.}
\end{biographies}
%

\end{document}